\documentclass[aps,preprint,floats,floatfix,epsfig,tightenlines]{revtex4}

\newcommand{\ba}{\begin{eqnarray}}
\newcommand{\ea}{\end{eqnarray}}
\def\ra{\rightarrow}
\def\eps{\epsilon}
\def\vareps{\varepsilon}

\usepackage{graphicx}

\begin{document}

%\tighten

\title{Limits on the Transient Ultra-High Energy Neutrino Flux from 
Gamma-Ray Bursts (GRB) Derived from RICE Data}

\author{D. Besson$^1$, S. Razzaque$^2$, J. Adams$^3$ and P. Harris$^3$}

\affiliation{\small $^1$Department of Physics and Astronomy, 
University of Kansas, Lawrence, KS 66045 \\ $^2$Department of
Astronomy and Astrophysics and Department of Physics, Pennsylvania
State University, University Park, PA 16802 \\ $^3$Department of
Physics, University of Canterbury, Christchurch, NZ}

%\vspace{1cm}

\begin{abstract}

We present limits on ultra-high energy (UHE; $E_\nu>10^{15}$ eV)
neutrino fluxes from gamma-ray bursts (GRBs), based on recently
presented data, limits, and simulations from the RICE experiment.  We
use data from five recorded transients with sufficient photon spectral
shape and redshift information to derive an expected neutrino flux,
assuming that the observed photons are linked to neutrino production
through pion decay via the well-known ``Waxman-Bahcall'' prescription.
Knowing the declination of the observed burst, as well as the RICE
sensitivity as a function of polar angle and the previously published
non-observation of any neutrino events allows an estimate of the
sensitivity to a given neutrino flux.  Although several orders of
magnitude weaker than the expected fluxes, our GRB neutrino flux
limits are nevertheless the first in the PeV--EeV energy regime.  For
completeness, we also provide a listing of other bursts, recorded at
times when the RICE experiment was active, but requiring some
assumptions regarding luminosity and redshift to permit estimates of
the neutrino flux.
\end{abstract}

%\date{}
\maketitle

%\tableofcontents

\section{Introduction}

\subsection{Gamma-ray Bursts}

Gamma-ray bursts (GRBs) are the most energetic explosions in the
Universe, releasing upwards of $\sim 10^{51}$ ergs of radiation energy
in $\gamma$-rays ($\sim 0.1$-$1$ MeV) within a few seconds (see
Refs. \cite{fm95, vkw00} for reviews on GRB observations). The Burst
and Transient Source Experiment (BATSE) on board the {\em Compton
Gamma-Ray Observatory} (CGRO) satellite detected bursts at the rate of
about 300 per year, in the period 1991-2001, isotropically distributed
over the sky. The observed duration of the bursts falls into two
categories, long bursts lasting for tens of seconds and short bursts
lasting for less than 2 seconds. The {\em Beppo}SAX, {\em High Energy
Transient Explorer} (HETE-2) and {\em SWIFT} satellites are observing
GRBs currently on a routine basis. However, the redshift measurement,
and hence the total energy estimate, is possible only for a small
fraction of the bursts as it requires identification of the GRB host
galaxies in optical wave bands.

The unsurpassed $\gamma$-ray luminosity naturally make GRBs candidate
sources of the observed ultra-high energy cosmic-rays \cite{v95,
w95a}, many with $\gtrsim 10^{18}$ eV.  High energy ($E_{\nu} \gtrsim
100$ GeV) neutrinos may then be created from the cosmic-rays, as in
beam-dump experiments, while propagating in the cosmic photon
backgrounds \cite{g66, s79, ess01} and/or at the particle acceleration
sites of the sources \cite{wb97a, wb98, wb00}. While diffuse flux
estimates are a useful method for calculating the total neutrino
energy injection into the universe by GRBs, individual source
detection is crucial for understanding the source production
processes, and in particular, for verifying whether point sources are
the sources of UHE cosmic rays. For transient sources such as GRBs,
individual luminous and bright bursts may dominate the observed
diffuse emission as well. A number of authors have calculated expected
neutrino production from individual long bursts (short bursts are
typically lower in energy output and hence less likely to be detected
individually in neutrino telescopes\cite{hh02, ghahr03, rmw03c, ahh03,
skc05, bshr05}).  When folded in with a given experimental acceptance,
such flux predictions can then be translated into expected
sensitivities or, in the case of no detection, upper limits.

\subsection{Radiowave Neutrino Detection}

\begin{figure}%[htpb]
\centerline{\includegraphics[width=8cm,angle=0]{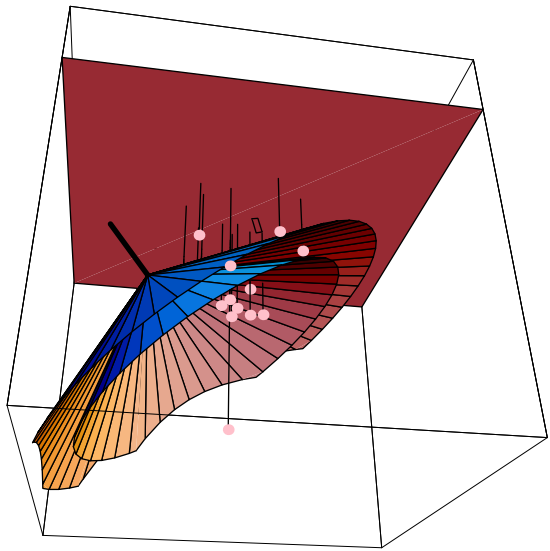}}
\caption{\it RICE concept. Shown is an incident neutrino interacting
$\sim$100m below the surface and illuminating the RICE radio antenna
array, indicated by spheres and drawn to scale. Horizontal extent is
approximately 400 m.}
\label{fig:concept}
\end{figure}

The RICE experiment has goals similar to the larger AMANDA and ICECUBE
experiments - both seek to measure UHE neutrinos by detection of
Cherenkov radiation produced by neutrino-nucleon interactions. Whereas
AMANDA/ICECUBE is optimized for detection of penetrating muons
resulting from $\nu_\mu+N\to\mu+N'$, RICE is designed to detect
compact electromagnetic and hadronic cascades initiated by charged or
neutral current neutrino interactions in a dense medium:
$\nu_l(/{\overline\nu_l}) + N\to l^\pm + N'$ or
$\nu_l(/{\overline\nu_l}) + N\to \nu_l(/{\overline\nu_l}) + N'$
respectively, with $l=e,\mu,\tau$. As the cascade evolves, a charge
excess develops as atomic electrons in the target medium are swept
into the forward-moving shower via Compton scattering and positrons
are depleted via annihilation, resulting in a net charge on the shower
front of $Q_{\rm tot}\sim E_se/4$; $E_s$ is the shower energy in GeV
\cite{ZHS, Alvarez-papers, SoebPRD, addendum, RomanJohn,
Shahid-hadronic, LPM, spencer04}.  Such cascades produce broadband
Cherenkov radiation -- for $\lambda^{\rm Cherenkov}_{\rm E-field}\gg
r_{\rm Moliere}$, the emitting region approximates a point charge of
magnitude $Q_{\rm tot}$ and therefore emits fully coherently;
fortuitously, the field attenuation length at such wavelengths in cold
polar ice ($\sim 1$ GHz) exceeds 1 km \cite{Barwick04}.  The quadratic
rise of the Cherenkov power with neutrino energy, combined with the
long attenuation length lead to the expectation that the effective
volume of radio-based detectors begin to exceed those of
photomultiplier-tube based detectors at a cross-over energy of $\sim$1
PeV \cite{Buford95}, leading to neutrino detector arrays consisting of
tens of buried radio antennas.  The RICE radio antenna detection
concept of the Cherenkov radiation produced by an in-ice weak current
event is depicted in Figure
\ref{fig:concept}.

In this paper we determine model-dependent RICE neutrino flux upper
limits for five GRBs, namely 020813, 020124, 050603, 050730 and 050908
for which we have extensive observational data including redshifts.
Using calculations published elsewhere of the expected radio-frequency
signal strength due to an electromagnetic shower
\cite{Alvarez-papers, SoebPRD}, the RICE hardware, reconstruction
software and simulation \cite{rice03a, rice03b, rice06}, and a
neutrino analysis based on data taken 1999-2005 \cite{rice06}, we can
translate an expected neutrino flux from the bursts recorded while
RICE was active into an expectation for the RICE neutrino yield.
These expectations are valid in an energy regime somewhat higher than
those typical of, e.g., previous AMANDA gamma-ray burst studies
\cite{ghahr03}.  Our calculation of expected neutrino fluxes from
these GRBs is the most novel aspect of this analysis, and is presented
in greater detail herein.

\section{Hardware}

The RICE experiment presently consists of a 20-channel array of dipole
radio receivers, scattered within a 200 m$\times$200 m$\times$200 m
cube, at 100-300 m depths.  The signal from each antenna is boosted by
a 36-dB in-ice amplifier, then carried by coaxial cable to the surface
observatory, where the signal is filtered (suppressing noise below 200
MHz), re-amplified (either 52- or 60-dB gain), and split - one copy is
fed into a CAMAC crate to form the event trigger; the other signal
copy is routed into one channel of an HP54542 digital
oscilloscope. Four antennas which register ``hits'' within a 1.25
microsecond window define an event trigger. ``Hits'' are defined as a
voltage excursion greater in magnitude than a discriminator threshold,
determined at the beginning of each run as that threshold which gives
approximately 90\% livetime. Deadtime is incurred by a full event
trigger readout (approximately 10 s of deadtime), the fast software
rejection of anthropogenic noise (approximately 10 ms of deadtime per
calculation), or the fast hardware rejection of anthropogenic noise
(1.2$\mu$sec per event).  Short-duration pulses broadcast from
under-ice transmitters provide the primary calibration signals and are
used to verify vertex reconstruction techniques.

Given the known experimental circuit gains and losses \cite{rice03a},
the effective volume $V_{eff}$ is calculated as a function of incident
$E_\nu$, as an exposure average of the detector configurations. The
most important variable is the global discriminator threshold, which
is adjusted to maintain an acceptable trigger rate under conditions of
varying environmental noise.

Additional details on the RICE shower search procedures are presented
elsewhere \cite{rice06}.  The most recent search has resulted in zero
neutrino candidates, yielding competitive limits on the incident
ultra-high energy neutrino flux at energies above $10^{17}$ eV.

\section{GRB model and neutrino flux calculation}

In the most widely accepted GRB model, known as the {\em fireball
shock model}, the prompt $\gamma$-rays are produced by collisions of
plasma material moving relativistically (with a bulk Lorentz factor
$\Gamma \gtrsim 100$) along a jet ({\em internal shocks}), i.e. a {\em
fireball}. Late time collisions of jetted material with an external
medium ({\em external shocks}) produce X-ray, UV and optical
radiation, collectively known as GRB afterglow (see Ref. \cite{m02,
p99} for reviews). A significant number of nucleons are expected to be
present in the jet along with leptons (electrons and positrons). The
probable mechanism(s) responsible for the observed photons is/are
synchrotron radiation or/and inverse Compton scattering by high energy
electrons.  These electrons are stochastically accelerated by
relativistic shocks via the well-known Fermi mechanism in the tangled
local magnetic field resulting in a power-law energy distribution.

Protons are expected to co-accelerate with electrons in the
shocks. High energy non-thermal neutrinos are expected to be produced
by photonuclear ($p\gamma$) interactions of protons with observed
prompt $\gamma$-rays \cite{wb97a, wb98} and afterglow photons
\cite{wb00} in these optically thin (for Thomson scattering)
environments. In this analysis, we put limits on the Waxman-Bahcall
GRB neutrino burst and neutrino afterglow flux models, which are based
on the standard fireball-shock model of GRBs. We do not consider in
the present work other neutrino flux models, e.g., (i) precursors from
GRB jets buried under the stellar surface \cite{mw01, rmw03b}; (ii)
GRB {\em blast-wave} models \cite{da03} in which both prompt and
afterglow emission arise from the external shocks; (iii) afterglow
from GRBs in dense stellar winds \cite{dl01}; (iv) the {\em
supranova} model \cite{gg03b, gg03c, rmw03a} in which a supernova type
remnant shell from the progenitor star is ejected prior to the GRB
event. \message{These models are either seldom used to interpret data
or there is a lack of evidence, especially from electromagnetic
observations, supporting them.}  Precursor neutrinos \cite{mw01,
rmw03b} below the RICE threshold ($E_{\nu} \gtrsim 10$ PeV) may be
detected by neutrino telescopes (e.g. IceCube) without any detectable
electromagnetic counter-part from their optically thick environment.

Below, we detail the prescription we have employed in estimating the
expected neutrino fluxes from a GRB with a well-measured $\gamma$-ray
flux, following closely the calculations of Waxman and Bahcall
(\cite{wb97a, wb98, wb00}, see also \cite{w01, rm-yellow}).

\subsection{GRB observed and model parameters}

The isotropic-equivalent $\gamma$-ray bolometric energy and luminosity
of a GRB at redshift $z$, given the measured bolometric $\gamma$-ray
fluence $S_{\gamma}$ (typically measured over one decade in energy
range) and duration $t_{90}$, may be calculated as
\ba
E_{\gamma,\rm iso} = 4\pi d_L^2 S_{\gamma}/(1+z) ~;~ L_{\gamma,\rm
iso} = 4\pi d_L^2 S_{\gamma}/t_{90},
\label{eq:bol-energy-luminosity}
\ea
where $d_L$ is the luminosity distance. The actual jetted energy is
$(4\pi/\Omega) E_{\gamma,\rm iso}$, where $\Omega$ is the jet opening
solid angle (typically $\sim 0.1$ radian). The $\gamma$-ray energy,
arising from shock-accelerated electron synchrotron radiation and/or
inverse Compton scattering, is assumed to be a fraction $\vareps_e <
1$ of the isotropic-equivalent kinetic energy, $E_{\rm kin, iso}$, of
the relativistic plasma material in the standard fireball-shock
model. The shock magnetic field shares another fraction $\vareps_B <
1$ of $E_{\rm kin, iso}$.

The luminosity distances may be calculated using the conversion
calculator provided by the NASA Extra galactic Database (NED
\cite{NEDref}) for a $\Lambda$CDM flat cosmology with $H_0 = 70$
km s$^{-1}$ Mpc$^{-1}$, $\Omega_{m}=0.3$ and $\Omega_{\Lambda} =0.7$
($1 ~{\rm Gpc} = 3.1\times 10^{27}$ cm). We have tabulated different
observed quantities in Table \ref{tab:grb-parameters} from 5
GRBs. The calculated values of $E_{\gamma,\rm iso}$ and
$L_{\gamma,\rm iso}$ are also tabulated for later use. The average
fractional RICE livetime over the duration of the burst is also
indicated in the last column.

The observed $\gamma$-ray spectrum from a GRB may be approximately
fitted with a broken power-law (Band fit \cite{b93}) as
\ba
\frac{dN_{\gamma}}{d\eps_{\gamma}} \propto \cases{ 
\eps_{\gamma}^{-\alpha} ~;~ \eps_{\gamma} < \eps_{\gamma,b} \cr
\eps_{\gamma}^{-\beta} ~;~ \eps_{\gamma} > \eps_{\gamma,b}. }
\label{eq:Band-fit}
\ea
The spectral indices and break energy $\eps_{\gamma,b}$ from
observations (where available) are listed in Table
\ref{tab:grb-parameters}. Values with asterisks have
been estimated using the phenomenological Ghirlanda relationship
\cite{ggf05} as
\ba
\eps_{\gamma,b} = \frac{300}{1+z} 
\left( \frac{E_{\gamma,\rm iso}}{10^{53}~{\rm ergs}} \right)^{0.56} 
~{\rm keV}.
\label{eq:Ghirlanda-relation}
\ea
Note, however, that the precise value of $\eps_{\gamma,b}$ is not
necessary to estimate the expected neutrino flux in the energy range
of our interest, as we calculate in the next subsection.

\begin{table}
\caption{\label{tab:grb-parameters} 
Observed and estimated GRB parameters}
\begin{ruledtabular}
\begin{tabular}{lcccccccccc} 
GRB & $S_\gamma$ & $t_{90}$ & $z$ & $d_L$ & $\eps_{\gamma,b}$ &
$\alpha$ & $\beta$ & $E_{\gamma,\rm iso}$ & $L_{\gamma,\rm iso}$ & 
RICE\\ & (ergs/cm$^2$) & (s) & & (Gpc) & (keV) & & & (ergs) & 
(ergs/s) & $<Live~T>$ 
\\ \hline
050908 & $5.1\times 10^{-7}$ & 20  & 3.35 & 29.02 & 21$^\ast$ 
       & - & 1.93 & $1.2\times 10^{52}$ & $2.6\times 10^{51}$ & 0.93 \\
050730 & $4.4\times 10^{-6}$ & 155 & 3.97 & 35.52 & 71$^\ast$ 
       & - & - & $1.3\times 10^{53}$ & $4.3\times 10^{51}$ & 0.75 \\
050603 & $3.4\times 10^{-5}$ & 10  & 2.82 & 23.59 & 289 
       & 0.79 & 2.15 & $5.9\times 10^{53}$ & $2.3\times 10^{53}$ & 0.92 \\
020124A & $6.1\times 10^{-6}$ & 46  & 3.20 & 27.47 & 87 
       & - & - & $1.3\times 10^{53}$ & $1.2\times 10^{52}$ & 0.83 \\
020813A & $8.4\times 10^{-5}$ & 89  & 1.25 & 8.71  & 142 
       & - & 2.15 & $3.4\times 10^{53}$ & $8.6\times 10^{51}$ & 0.98
\end{tabular} 
\end{ruledtabular}
\end{table}

\subsection{Burst neutrinos} 

The condition for resonant $p\gamma$ interaction at $\Delta^+$ energy
is $\eps'_{p}\eps'_{\gamma} \simeq 0.3 ~{\rm GeV}^2$ in the local rest
frame of the plasma [We denote primed (unprimed) variables in the
plasma rest (lab) frame. The obesrved variables are redshift-corrected
lab variables]. On average, the proton loses a fraction $<\!\!
x_{p\ra \Delta^+}\!\!> \simeq 0.2$ of its energy in this
interaction. The secondary $\Delta^+$, produced by a $p\gamma$
interaction decays as either $\Delta^+ \ra n\pi^+$ or $\Delta^+ \ra
p\pi^0$.\footnote{We follow the original Waxman-Bahcall estimate,
which took the ratio of $\pi^+:\pi^0$=1:1. The true ratio (from
isospin) is 1:2.} The pions further decay as $\pi^+ \ra \mu^+
\nu_{\mu} \ra e^+ \nu_e {\bar \nu}_{\mu} \nu_{\mu}$ or $\pi^0 \ra 
\gamma\gamma$. In what follows, we assume equipartition of energy 
among the 4 final state leptons from $\pi^+$-decay. The observed
neutrino break energy corresponding to the $\Delta^+$ resonance is
then
\ba
\eps_{\nu, b1} = \frac{0.015 \Gamma_i^2}{(1+z)^2} 
\left( \frac{\eps_{\gamma, b}}{\rm GeV} \right)^{-1} ~{\rm GeV},
\label{eq:burst-nu-break}
\ea
where $\eps_{\gamma,b}$ is the break energy in the $\gamma$-ray
spectrum [or the peak energy in the $\eps_{\gamma}^2
(dN_{\gamma}/d\eps_{\gamma})$ energy spectrum]. The synchrotron break
energy (above which pions lose significant energy by synchrotron
radiation before decaying to neutrinos) in the neutrino spectrum is
\ba
\eps_{\nu, sb} = \frac{10^{11}\Gamma_i}{4(1+z)} 
\left( \frac{B'_i}{\rm G} \right)^{-1} ~{\rm GeV}, 
\label{eq:sync-break}
\ea
where the magnetic field in the GRB fireball is given by
\ba
B'_i=\sqrt{\frac{2\vareps_B L_{\gamma,\rm iso}}{\vareps_e r_i^2
\Gamma_i^2 c}} = 5\times 10^4 
\left( \frac{L_{\gamma,\rm iso}}{10^{52} ~{\rm ergs/s}} \right)^{1/2}
\left( \frac{\Gamma_i}{300} \right)^{-3}
\left( \frac{t_v}{0.01~{\rm s}} \right)^{-1} ~{\rm G}.
\label{eq:internal-B-field}
\ea
The typically unknown parameters are $\vareps_e \approx \vareps_B
\approx 0.1$; the Lorentz factor of the GRB fireball is 
$\Gamma_i \approx 300$ from various fits. The radius of the fireball
is $r_{i}\simeq 2 \Gamma_i^2 ct_v$, which varies over a large range
depending on $\Gamma_i$ and the variability time scale (typically,
0.001 s $\le t_v \le 1$ s).

To calculate the expected neutrino flux from the GRB fireball, we
assume a shock-accelerated proton luminosity in the fireball $L_{p,\rm
iso} \simeq L_{\gamma,\rm iso}$ with energy distributed as a
power-law: $dN_p/d\eps_p \propto \eps_p^{-2}$ up to the maximum energy
$\lesssim 10^{20}$ eV. The total proton energy, equal per decade in
energy, is then $\sim 10E_{\gamma,\rm iso}$ the $\gamma$-ray energy;
roughly consistent with the assumed equipartition fraction $\vareps_e
\sim 0.1$ of $E_{\rm kin, iso}$. Due to interactions of these protons
with the observed $\gamma$-rays the corresponding neutrino energy flux
from the GRB prompt phase, following Eqs. (\ref{eq:Band-fit}),
(\ref{eq:burst-nu-break}) and (\ref{eq:sync-break}), is given by:
\ba
\eps_{\nu}^2 \Phi_{\nu}^s = \frac{1}{2}\frac{f_{\pi}}{4} 
\frac{S_{\gamma}}{t_{90}} \times \cases{ 
(\eps_{\nu}/\eps_{\nu,b1})^{\beta-1} ~;~ \eps_{\nu} < \eps_{\nu,b1}
\cr (\eps_{\nu}/\eps_{\nu,b1})^{\alpha-1} ~;~
\eps_{\nu,sb} \ge \eps_{\nu} \ge \eps_{\nu,b1} \cr
(\eps_{\nu,sb}/\eps_{\nu,b1})^{\alpha-1} 
(\eps_{\nu}/\eps_{\nu,sb})^{-2} ~;~ \eps_{\nu} > \eps_{\nu,sb}, }
\label{eq:burst-flux}
\ea
for each neutrino flavor $\nu_{\mu}$, ${\bar \nu}_{\mu}$ and $\nu_e$
at the source. The factors $1/2$ and $1/4$ arise due to the WB
assumption of equal decay probabilities: $\Delta^+ \to \pi^+/\pi^0$
and energy equipartion among the $\pi^+$ decay products,
respectively. Here $f_\pi$ is the conversion efficiency of shock
accelerated protons to pions at the $\Delta^+$ resonance in the GRB
fireball. For optically thin sources, such as GRBs, $f_{\pi} \le
0.2$. The detailed calculation of $f_\pi$, which involves many unknown
variables, is not relevant here. Moreover, the internal shocks (and
thus $\gamma$-ray emission) occur over a wide range of $r_i$ which in
turn affect $f_\pi$. We assume a fixed value $f_{\pi} =0.2$ for our
calculation which corresponds to a $p\gamma \to \Delta^+$ optical
depth of the order of unity.  The prefactor in the expression for flux
[Eq. (\ref{eq:burst-flux})] can be written as
\ba
{\cal A} \equiv \frac{f_{\pi}}{8} \frac{S_{\gamma}}{t_{90}} =
1.56\times 10^{-6} \left( \frac{f_{\pi}}{0.2} \right)
\left( \frac{S_{\gamma}}{10^{-6}~{\rm ergs/cm^2}} \right)
\left( \frac{t_{90}}{10~{\rm s}} \right)^{-1} 
~{\rm GeV ~cm^{-2} ~s^{-1},}
\label{eq:burst-normalization}
\ea
and the $\gamma$-ray spectral indices are $\alpha = 1$, $\beta =2$ to
good approximation ($1~{\rm erg} = 624 ~{\rm GeV}$). After
oscillations in vacuum, the neutrino fluxes for different flavors are
${\Phi_{\nu_e + {\bar \nu}_e}} = {\Phi_{\nu_{\mu} + {\bar \nu}_{\mu}}}
= {\Phi_{\nu_{\tau} + {\bar \nu}_{\tau}}} = \Phi_{\nu}^s$, each with a
duration of $t_{\rm burst}=t_{90}$. We have listed values for these
parameters in Table \ref{tab:burst-parameters} for 5 bursts.

\begin{table}
\caption{\label{tab:burst-parameters} Burst neutrino flux parameters}
\begin{ruledtabular}
\begin{tabular}{lcccc} 
GRB & ${\cal A}$ (GeV cm$^{-2}$ s$^{-1}$) & $\eps_{\nu, sb}$ (GeV) & 
$\eps_{\nu, b1}$ (GeV) & $t_{\rm burst} (s)$ \\ \hline
050908 & $4.0\times 10^{-7}$ & $9.5\times 10^7$ & $3.4\times 10^6$ & 20 \\
050730 & $4.4\times 10^{-7}$ & $6.4\times 10^7$ & $7.7\times 10^5$ & 155 \\
050603 & $5.3\times 10^{-5}$ & $1.2\times 10^7$ & $3.2\times 10^5$ & 10 \\
020124A & $2.1\times 10^{-6}$ & $4.6\times 10^7$ & $8.8\times 10^5$ & 46 \\
020813A & $1.5\times 10^{-5}$ & $1.0\times 10^8$ & $1.9\times 10^6$ & 89 \\
\end{tabular}
\end{ruledtabular}
\end{table}

\subsection{Afterglow neutrinos}

The GRB afterglow, arising from external forward and reverse shocks,
takes place at a radius $r_e = (3 E_{\rm kin, iso}/[4\pi n_{\rm ex}
m_p c^2 \Gamma_i^2])^{1/3}$, which is much larger than $r_i$. Here we
assume the total jet kinetic energy $E_{\rm kin, iso} = E_{\gamma,\rm
iso}/\vareps_e \simeq 10 E_{\gamma,\rm iso}$ throughout our
calculation and $n_{\rm ex} = 1$ atom cm$^{-3}$ is a typical
interstellar medium (ISM, dominantly neutral hydrogen) density. The
bulk Lorentz factor of the reverse-shocked plasma shell at the time
$t=t_{90}$ when the reverse shock has crossed the shell is
\ba \Gamma_e = \frac{1}{4} \left[ \frac{17 E_{\rm kin, iso}} 
{\pi n_{\rm ex} m_p c^5 t_{90}^3} \right]^{1/8} \approx 195
\left( \frac{E_{\rm kin, iso}}{10^{54}~{\rm ergs}} \right)^{1/8} 
\left( \frac{t_{90}}{10~{\rm s}} \right)^{-3/8}
\left( \frac{n_{\rm ex}}{1~{\rm cm}^{-3}} \right)^{-1/8}.
\label{BM-gamma}
\ea

The reverse shock magnetic field in the comoving plasma frame is
$B'_e=(8\pi \vareps_B n_{\rm ex} m_pc^2)^{1/2}$ and we set
$\vareps_B=0.1$ as in the internal shocks. The minimum Lorentz factor
of the reverse shock-accelerated electrons in the un-shocked plasma
frame is $\gamma'_{e,\rm min} \simeq \vareps_e (m_p/m_e)
(\Gamma_i/\Gamma_e)$ and the peak electron synchrotron photon energy
is $\eps'_{\gamma, m} = \hbar c \Gamma_e (3\gamma_{e,\rm min}^{'2} e
B'_e) /(2m_e c^2)$. However the characteristic shock-accelerated
electron Lorentz factor, from equating the synchrotron cooling time to
the dynamic time: $t_{\rm dyn} = r_e/4\Gamma_e^2c$, is $\gamma'_{e,c}
= 6\pi m_e c /(\sigma_{\rm Th} B_e^{'2} \Gamma_e^4 t_{\rm dyn})$.  The
corresponding peak synchrotron photon energy radiated by these
characteristic electrons in the same $B'_e$ field is then
\ba 
\eps'_{\gamma, c} = \hbar c \Gamma_e 
\frac{3\gamma_{e,c}^{'2} e B'_e} 
{2m_e c^2} \approx 0.4
\left( \frac{E_{\rm kin, iso}}{10^{54}~{\rm ergs}} \right)^{-25/24}
\left( \frac{\Gamma_i}{300} \right)^{4/3}
\left( \frac{t_{90}}{10~{\rm s}} \right)^{9/8}
\left( \frac{n_{\rm ex}}{1~{\rm cm}^{-3}} \right)^{-11/24} ~{\rm eV}.
\label{eq:charct-photon-E}
\ea
The number density of these photons in the reverse shock region is
approximately given by the total synchrotron radiation power
(typically at $\eps'_{\gamma, m}$) by all electrons scaled at
$\eps'_{\gamma, c}$ and then divided by $\eps'_{\gamma, c}$ as
\ba 
n'_{\gamma, c} &\simeq & \frac{E_{\rm kin, iso}}{8\pi^2 \hbar c r_e^2}
\left( \frac{\Gamma_e}{\Gamma_i} \right) \frac{e^3 B'_e}{m_ec^2 m_pc^2} 
\left( \frac{\eps'_{\gamma, m}}{\eps'_{\gamma, c}} \right)^{1/2} 
\nonumber \\ &\approx & 6.8\times 10^{12}
\left( \frac{E_{\rm kin, iso}}{10^{54}~{\rm ergs}} \right)^{11/12}
\left( \frac{\Gamma_i}{300} \right)^{2/3}
\left( \frac{t_{90}}{10~{\rm s}} \right)^{-3/4}
\left( \frac{n_{\rm ex}}{1~{\rm cm}^{-3}} \right)^{19/12}
~{\rm cm}^{-3}.
\label{eq:charct-photon-dens}
\ea
The corresponding photon number spectrum is given by
\ba 
\eps'_{\gamma} \frac{dN'_{\gamma}}{d\eps'_{\gamma}} \simeq
n'_{\gamma, c} \times \cases{ (\eps'_{\gamma}/\eps'_{\gamma,
c})^{-1/2} ~;~ \eps'_{\gamma, m}< \eps'_{\gamma}< \eps'_{\gamma,
c} \cr (\eps'_{\gamma}/\eps'_{\gamma, c})^{-1} ~;~
\eps'_{\gamma}>\eps'_{\gamma, c}.  }
\label{eq:ext-photon-spect}
\ea

Note that, one needs to explicitly calculate the proton to pion
conversion efficiency factor $f_\pi$ in the afterglow phase ($r_e >
r_i$) as it may be lower than 0.2, contrary to the prompt (internal
shocks) phase. We calculate $f_\pi$ for reverse shock-accelerated
protons to interact with photons of energy $\eps'_{\gamma, c}$
[Eq. (\ref{eq:charct-photon-E})] at the $\Delta^+$ resonance as
\ba 
f_{\pi} ={\rm min} (1, \tau'_{p\gamma \ra \Delta^+}) <\!\! x_{p\ra
\Delta^+} \!\!> \approx 0.2
\left( \frac{E_{\rm kin, iso}}{10^{54}~{\rm ergs}} \right)^{33/24}
\left( \frac{t_{90}}{10~{\rm s}} \right)^{-9/8}
\left( \frac{n_{\rm ex}}{1~{\rm cm}^{-3}} \right)^{9/8},
\label{eq:ext-fpi}
\ea
where $\tau'_{p\gamma \ra \Delta^+} \simeq \sigma_{p\gamma \ra
\Delta^+} (r_e/\Gamma_e) n'_{\gamma,c}$ is the optical depth using
Eq. (\ref{eq:charct-photon-dens}) and $\sigma_{p\gamma \ra
\Delta^+} = 5\times 10^{-28}$ cm$^2$ is the $p\gamma$ cross-section at
the $\Delta^+$ resonance. The neutrino break energy for the
characteristic photon energy [Eq. (\ref{eq:charct-photon-E})] is
\ba
\eps_{\nu, b2} = \frac{0.015 \Gamma_e}{(1+z)} 
\left( \frac{\eps'_{\gamma, c}}{\rm GeV} \right)^{-1} ~{\rm GeV}.
\label{eq:ext-nu-break}
\ea
The maximum shock-accelerated proton energy can be found by equating
the shock-acceleration time to the shorter of the synchrotron cooling
time (most efficient energy loss channel) and the dynamic time. The
corresponding maximum neutrino energy is
\ba
\eps_{\nu,\rm max} = \frac{<\!\! x_{p\ra \Delta^+} \!\!> } {4(1+z)}
(e\Gamma_e B'_e r_e)
\label{eq:ext-maxnu-E}
\ea
in the afterglow phase. The resulting afterglow neutrino spectrum,
following Eqs. (\ref{eq:ext-photon-spect}), (\ref{eq:ext-fpi}) and
(\ref{eq:ext-nu-break}), is
\ba
\eps_{\nu}^2 \Phi_{\nu}^s = \frac{f_{\pi}}{8} 
\frac{S_{\gamma}}{t_{90}} \times \cases{ 
(\eps_{\nu}/\eps_{\nu,b2}) ~;~ \eps_{\nu} < \eps_{\nu,b2}
\cr (\eps_{\nu}/\eps_{\nu,b2})^{1/2} ~;~ 
\eps_{\nu,\rm max} \ge \eps_{\nu} \ge \eps_{\nu,b2}, }
\label{eq:aftglow-flux}
\ea
for each neutrino flavor $\nu_{\mu}$, ${\bar \nu}_{\mu}$ and $\nu_e$
at the source. Again, the flavor ratios on Earth, after vacuum
oscillation, would be $\Phi_{\nu_e + {\bar \nu}_e} =
\Phi_{\nu_{\mu} + {\bar \nu}_{\mu}} = 
\Phi_{\nu_{\tau} + {\bar \nu}_{\tau}} = 
\Phi_{\nu}^s$, each with a 
duration of $t_{\rm glow} = t_{90}$. The parameters $f_{\pi}$,
$\eps_{\nu,b2}$ and $\eps_{\nu,\rm max}$ are listed in 
Table \ref{tab:afterglow-parameters}
for each GRB. The pre-factor in the neutrino spectrum ${\cal B} =
(f_{\pi}/8) (S_{\gamma}/t_{90})$, for different GRBs, is also listed
in Table \ref{tab:afterglow-parameters}.

\begin{table}
\caption{\label{tab:afterglow-parameters} 
Afterglow neutrino flux parameters}
\begin{ruledtabular}
\begin{tabular}{lccccc} 
GRB & $f_{\pi}$ & ${\cal B}$ (GeV/cm$^2$/s) & $\eps_{\nu, b2}$ (GeV) &
$\eps_{\nu,\rm max}$ (GeV) & $t_{\rm glow} (s)$ \\ \hline
050908 & 0.03 & $6.9\times 10^{-8}$ & $1.5\times 10^8$ & 
$1.4\times 10^8$ & 20 \\
050730 & 0.2  & $4.4\times 10^{-7}$ & $2.5\times 10^8$ & 
$2.8\times 10^8$ & 155 \\
050603 & 0.2   & $5.3\times 10^{-5}$ & $1.0\times 10^{10}$ & 
$6.0\times 10^8$ & 10 \\
020124A & 0.2  & $2.1\times 10^{-6}$ & $7.4\times 10^8$ & 
$3.3\times 10^8$ & 46 \\
020813A & 0.2   & $1.5\times 10^{-5}$ & $2.0\times 10^9$ & 
$8.4\times 10^8$ & 89
\end{tabular}
\end{ruledtabular}
\end{table}

\begin{figure}%[htpb]
\centerline{\includegraphics[width=8cm,angle=0]{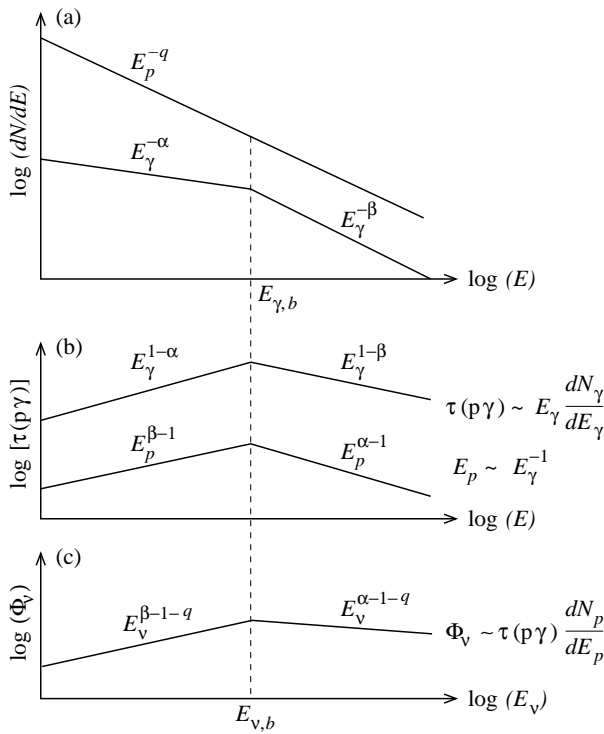}}
\caption{\it 
Cartoon of the neutrino energy spectrum expected from a GRB having an
observed $\gamma$-ray spectrum. (a) Shock accelerated proton spectrum:
$\propto \eps_p^{-q}$, and broken power-law fit to the $\gamma$-ray
spectrum: $\propto \eps_{\gamma}^{-\alpha}$ for
$\eps_{\gamma}<\eps_{\gamma,b}$ and
$\eps_{\gamma}>\eps_{\gamma,b}$. (b) Opacity for $p\gamma \to
\Delta^+$ resonant interaction satisfying the condition $\eps_p
\eps_{\gamma} = 0.3 \Gamma^2$ GeV$^2$. Note that the optical depth 
changes its shape when expressed as a function of $\eps_p$ (lower
curve) rather than $\eps_{\gamma}$ (upper curve) to satisfy the
resonance condition $\eps_p\eps_{\gamma} =$ const. (c) The neutrino
spectrum follows the proton spectrum modulo the shape of the optical
depth. At high energy (not shown here) above the synchrotron break
energy [Eq. (5)], the neutrino spectrum would steepen by a factor of 2
in the spectral index.}
\label{fig:grb2nu}
\end{figure}

The formalism outlined above is schematically presented in Figure
\ref{fig:grb2nu}, which qualitatively indicates the relationship
between the neutrino and the photon spectra.

\subsection{GRBs with incomplete data}

We note $\sim$100 GRBs (listed in the Appendix) having insufficient
electromagnetic data recorded while RICE was live. The break energy
($\eps_{\gamma,b}$) and spectral indices ($\alpha$ and $\beta$) of the
$\gamma$-ray spectrum may be estimated [e.g., $\eps_{\gamma,b}$ from
Eq. (\ref{eq:Ghirlanda-relation})] and/or sampled from their observed
distributions. The relative error using the estimated spectral
information is small and does not largely affect the neutrino flux
predictions in the energy range where RICE is sensitive.  However, the
most important piece of information that is lacking is the redshift
measurement. The luminosity and hence energy of a burst [see
Eq. (\ref{eq:bol-energy-luminosity})] depends on the redshift or its
luminosity distance. These in turn affect the model neutrino flux
calculation. Various methods have been proposed to indirectly measure
the redshift of a burst, e.g., from the time lag between the hard and
soft $\gamma$-ray arrival from a burst \cite{norris,bonnel}; using an
empirical relation between the observed variability and the luminosity
\cite{Reichart,Fenimore}; using observed correlation between $\gamma$-ray 
total and peak energy \cite{Amati02}; etc. For now, we have focused on
those bursts for which reliable redshift data exist.

\section{Search for Gamma-Ray Burst coincidences}

Software algorithms used to winnow the RICE data set are presented
elsewhere \cite{rice03a, rice03b, rice06}.  Typically, antenna `hits'
must exceed a threshold corresponding to approximately
$5.5\sigma_{rms}$, where $\sigma_{rms}$ is the rms noise in the
recorded waveform for a given antenna channel.  In cases for which the
event time is known $a~priori$, one can employ more sophisticated,
albeit more time-consuming event reconstruction techniques. These
include use of a `matched filter' to search for lower-amplitude
signals, thereby enhancing the effective volume. Although transients
such as GRBs lend themselves to such approaches, we have not attempted
such an analysis and instead rely on the results of previous RICE
analyses for sensitivity and efficiency estimates.

We select gamma-ray bursts which were recorded within a $\pm$1000 s
overlap time window during which the RICE experiment was active
\cite{hitsfiles}.  Only GRBs which have declination angles less than
zero and have therefore been localized to the southern hemisphere are
considered, given the opacity of the earth to neutrinos at ultra-high
energies.  The recorded RICE experimental discriminator threshold
allows a determination of the effective volume specifically for the
run during which the gamma-ray burst was observed.  The record of
triggers, and software surface-noise vetoes over the 2000 second
coincidence window allows an estimate of the average livetime fraction
over that period. Relative to the published RICE effective volume and
efficiency \cite{rice06}, our ultimate sensitivity is degraded by this
calculated average livetime fraction.

\begin{figure}%[htpb]
\centerline{\includegraphics[width=12cm,angle=0]{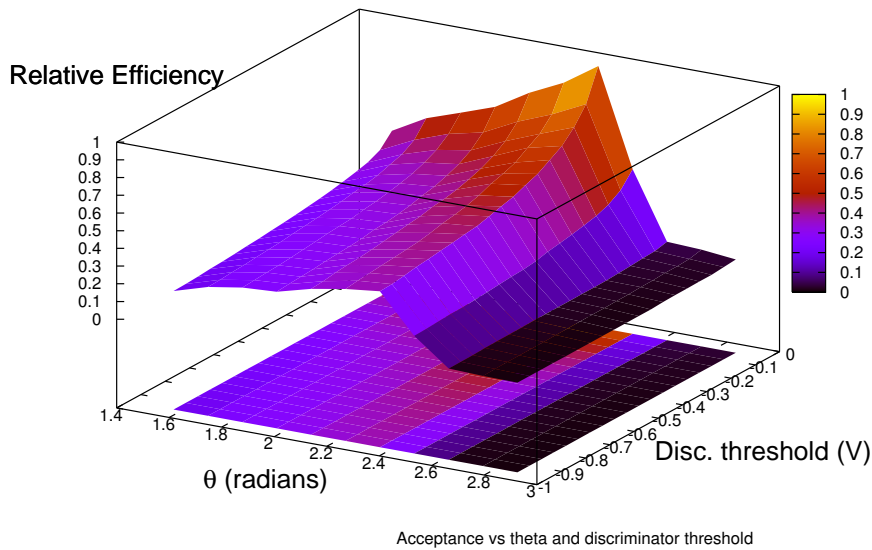}}
\caption{\it Relative dependence of $V_{eff}$ on nadir angle (i.e., 
$\pi/2$+declination), and discriminator threshold, based on
simulations of the RICE detector.  $\pi$/2 corresponds to flux
incident along the horizon; $\pi$ corresponds to flux incident from
zenith.}
\label{fig:vefftheta1}
\end{figure}

\begin{figure}%[htpb]
\centerline{\includegraphics[width=12cm,angle=0]{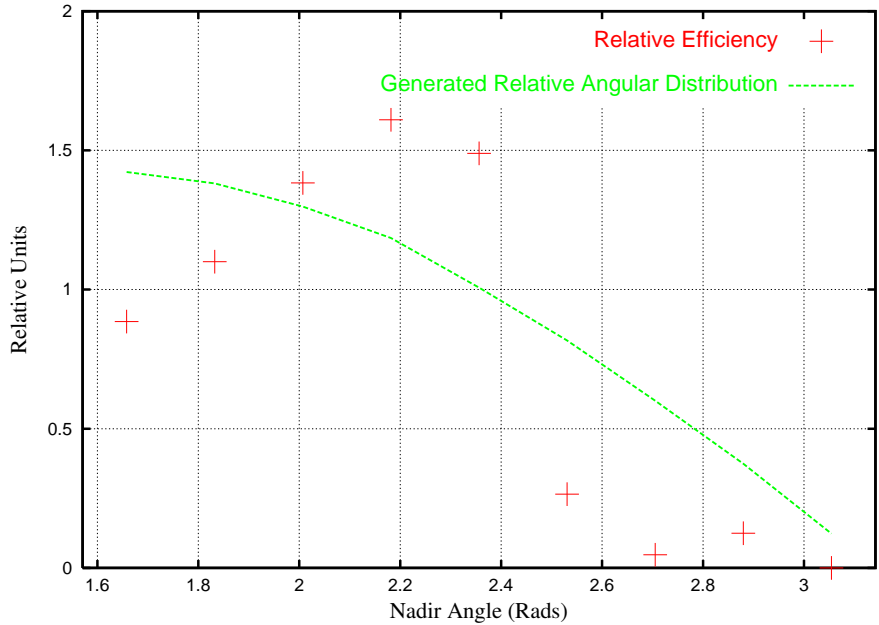}}
\caption{\it Dependence of $V_{eff}$ on nadir angle, integrated over
energy. Also shown is the distribution of the expected incident
flux. The overall efficiency is therefore the properly weighted
average of the polar angle dependence.}
\label{fig:vefftheta2}
\end{figure}

Relative to the effective volume integrated over angle \cite{rice06},
appropriate for an expected isotropic (i.e., flat in both
$dN/d(\cos\theta))$ and $dN/d\phi$) diffuse neutrino flux, we must
determine the acceptance for point sources as a function of
declination and azimuth. Our simulations indicate that the acceptance
is constant (to within $\sim$10\%) in azimuth and that the dominant
angular dependence is on the polar angle of the point source.  Since
the RICE array is shallow, the acceptance for neutrinos incident from
the zenith (directly overhead) is nearly zero. Maximal acceptance is
achieved for neutrinos incident closer to the horizon interacting in
the 2.5 km of ice below the array which illuminate the array with a
relative angle between the neutrino momentum vector and the vector to
the array close to the Cherenkov angle.  Figure \ref{fig:vefftheta1}
shows the relative effective volume as a function of the discriminator
threshold and the nadir angle of the incident neutrino. Given the fact
that the angular dependence is more important than the energy
dependence, Figure \ref{fig:vefftheta2} shows the angular dependence
of the effective volume, integrated over energy.  For each GRB, this
graph provides a correction relative to the previously-published
$V_{eff}$ \cite{rice06} using the tabulated declination angle for each
analyzed GRB.

\begin{figure}
\centerline{\includegraphics[width=12cm,angle=-90]{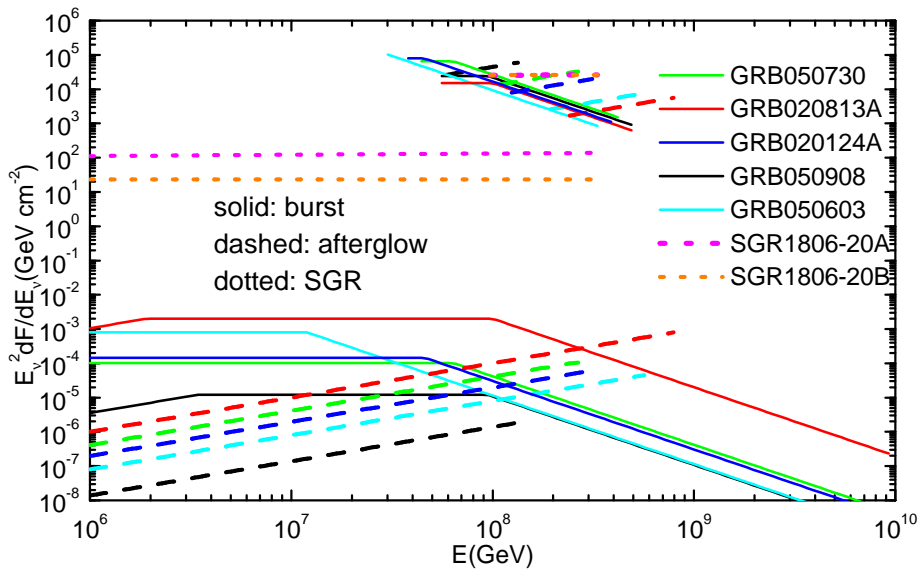}}
\caption{\it Upper limit, derived from five indicated GRBs and for SGR
1806-20, assuming a 
1:1:1 isoflavor mix at the detector. Lower set of lines indicate
predicted GRB and SGR 1860-20 neutrino fluxes as described in 
the text; higher set of
lines indicate RICE upper limits using previously published effective
volumes and the expected angular dependence of the acceptance.}
\label{fig:ricelimgrbpt}
\end{figure}

The non-observation of any in-ice shower candidates, coupled with the
tabulated RICE neutrino-finding efficiency, allows an estimate of the
expected event yield for GRB neutrino fluxes calculated as above.  Our
derived upper limits, following the prescription for relating a flux
model to an upper limit as outlined in the previous RICE neutrino flux
study \cite{rice06}, are presented in Figure
\ref{fig:ricelimgrbpt}. We also list the limit values for each GRB in
Table \ref{tab:limits} and their ratios to the predicted flux values
at a given neutrino energy. Simple linear scaling then may be used to
get limit values at another energy.

\begin{table}
\caption{\label{tab:limits} 
RICE limits and ratios to the predicted burst and afterglow 
fluxes at a given energy}
\begin{ruledtabular}
\begin{tabular}{l|ccc|ccc} 
GRB & & Burst & & & Afterglow & \\ 
    & $\eps_{\nu}$ (GeV) & Limit (GeV/cm$^2$) & Ratio 
    & $\eps_{\nu}$ (GeV) & Limit (GeV/cm$^2$) & Ratio \\
\hline
050908 & $1.0\times 10^{8}$ & 20152.48 & $2.01\times 10^{9}$ 
       & $1.0\times 10^{8}$ & 44487.72 & $3.21\times 10^{10}$ \\
050730 & $1.0\times 10^{8}$ & 24624.48 & $6.35\times 10^{8}$ 
       & $2.0\times 10^{8}$ & 25037.32 & $3.04\times 10^{8}$ \\
050603 & $1.0\times 10^{8}$ & 8613.60  & $8.13\times 10^{8}$ 
       & $2.0\times 10^{8}$ & 2739.27  & $1.68\times 10^{8}$ \\
020124A & $1.0\times 10^{8}$ & 15436.97 & $5.44\times 10^{8}$ 
        & $2.0\times 10^{8}$ & 12874.04 & $3.24\times 10^{8}$ \\
020813A & $1.0\times 10^{8}$ & 13975.23 & $7.55\times 10^{6}$ 
        & $3.0\times 10^{8}$ & 2139.85 & $7.05\times 10^{6}$
\end{tabular} 
\end{ruledtabular}
\end{table}

\begin{figure}
\centerline{\includegraphics[width=12cm,angle=0]{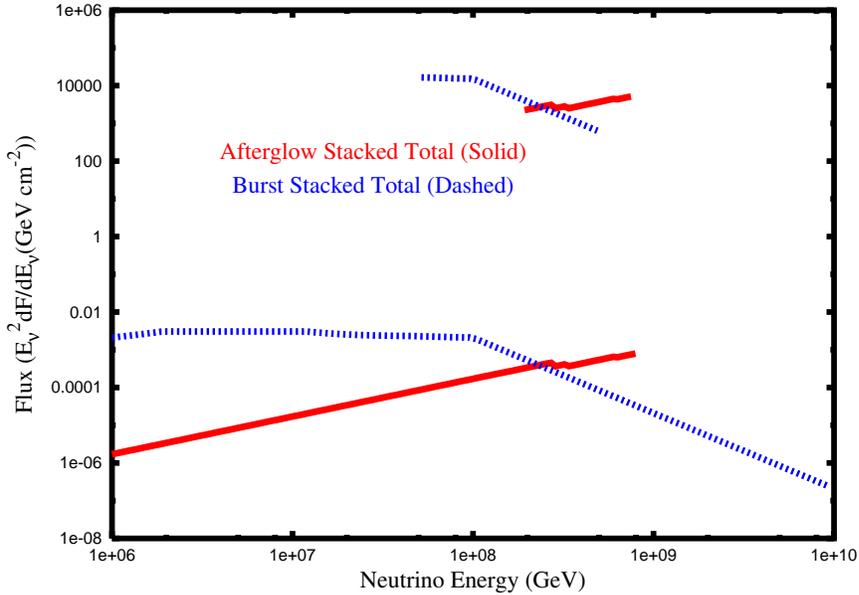}}
\caption{\it Upper limit on the total stacked flux from the 
five GRBs, 020813, 020124, 050603, 050730 and 050908, assuming a 1:1:1
isoflavor mix at the detector. The lower pair of lines indicate the
predicted total neutrino flux from the five GRBs for the burst and
afterglow models for neutrino production; the higher set of lines
indicate RICE upper limits for the total neutrino fluxes using
previously published effective volumes and the expected angular
dependence of the acceptance.}
\label{fig:stackedlimit}
\end{figure}

An alternative approach to setting limits on the neutrino flux from
individual GRBs is to use a source stacking method. These methods have
been applied in gamma-ray astronomy and were discussed recently in the
context of neutrinos from various classes of active galactic nuclei
(AGNs) \cite{andreas}. The total neutrino flux from all five GRBs in
our sample, for the burst and afterglow models described, is displayed
in Figure \ref{fig:stackedlimit}. We derive upper limits on these
total fluxes by following the same prescription as for the single
burst analysis, using the sum of the number of events expected from
each individual GRB.

A class of low luminosity, low energy (dominantly in the X-ray bands)
bursts are known as X-ray flashes (XRFs). The typical break energy
$\eps_{\gamma,b} \sim$ 1-10 keV implies that protons need to be
accelerated to $\gtrsim 10^{18}$ eV to produce neutrinos at these
sources which is unlikely because of their low luminosity. Moreover,
the proton to pion conversion efficiency $f_\pi$ would be much lower
in this case. There were 3 XRFs recorded while RICE was active namely
GRBs 020903A, 030429A and 030723A (see Appendix) with bolometric
fluences $S_{\gamma}$ one to two orders of magnitude lower than
typical long GRBs listed in Table \ref{tab:grb-parameters}.

\subsection{Other Transient Source Searches}

Soft Gamma-ray Repeaters (SGRs) represent one class of extreme X-ray
pulsars repeatedly emitting $\sim 0.1$ s bursts of soft
$\gamma$-rays. SGRs are highly magnetized ($\sim 10^{15}$ G) neutron
stars, often called `magnetars'. Four SGRs have been identified in the
Milky Way (1806-20, 1627-41, 1900+14, and 1801-23) and one has been
identified in the Large Magellanic Clouds (0526-66). Giant flares from
an SGR, hypothesized to originate from a sudden shift of magnetic
field, occur with periods of $\sim$ 10 years. In particular, the giant
flare of SGR 1806-20 on 27 December 2004 at 21:30:26 UT was the
brightest cosmic transient observed to date with a flux $\sim 10$ ergs
s$^{-1}$ cm$^{-2}$ lasting for $\sim 0.1$ s. We note that this flux is
$\sim 10^4$ times larger than the most luminous GRBs ever
detected. Detected radio afterglow of SGRs imply the presence of
relativistic jets, similar to GRBs. If high energy neutrinos were
emitted with comparable energy emitted in $\gamma$-rays, the 27
December 2004 giant flare of SGR 1806-20 might have been detected by
operating neutrino telescopes, as predicted by several authors
\cite{irkm05, hlm05}.

We investigated the possibility of a coincidence with the 27 Dec 04
SGR 1806-20 giant flare. Although the RICE experiment was active
during that time, we did not observe a trigger at the time of the SGR
flare.  We did record two forced triggers (``unbiased'' events) at
21:25:33 UT and 21:35:33 UT, respectively, which bracket the flare
time by almost exactly five minutes.  A channel-by-channel comparison
of the time-domain and frequency-domain waveforms for these events
reveals no obvious differences between the data recorded before-flare
vs. after-flare. Given no detection we then put limits on two model
fluxes
\ba
\eps_{\nu}^2 \Phi_{\nu, pp}^{\rm s} &=& 330+30~{\rm ln} 
\left( \frac{\eps_{\nu}}{\rm GeV} \right) 
~{\rm GeV}~{\rm cm}^{-2} {\rm s}^{-1} ~{\rm (Model~A)} \nonumber \\
\eps_{\nu}^2 \Phi_{\nu, p\gamma}^{\rm s} &=& 156 
~{\rm GeV}~{\rm cm}^{-2} {\rm s}^{-1} ~{\rm (Model~B)}
\ea
where Model A corresponds to the neutrino flux originating from $pp$
interactions in Ref. \cite{irkm05} and Model B is a generic $p\gamma$
flux, similar to the prediction in Ref. \cite{hlm05}, assuming total
proton energy is equal to the observed $\gamma$-ray energy. The
maximum neutrino energy in both cases is $3.3\times 10^8$ GeV
\cite{irkm05}. The flux predictions and the limits are plotted in
Fig. \ref{fig:ricelimgrbpt}. The limits are $191$ and $550$ times
higher than the flux predictions for Models A and B respectively.

\section{Summary}

Selecting GRB events recorded while the RICE experiment was actively
taking data, and using previously presented results on the RICE
analysis procedures and simulations, we have estimated the expected
RICE sensitivity to such transients. Based on the lack of observed
coincidences, we set limits on the UHE neutrino flux for several
candidates. Although the limits presented are well above model
predictions, they do, nevertheless, probe possible neutrino fluxes
from GRBs in an energy regime considerably higher than those previous
studied. Future improvements in sensitivity will be realized by
enlarging the scale of the radio detector array, as well as improved
individual GRB parameter measurements.

\section{Acknowledgments}

This work was supported by the National Science Foundation Office of
Polar Programs and the Department of Energy.  We gratefully
acknowledge the generous logistical support of the AMANDA and SPASE
Collaborations (without whom this work would not have been possible),
the National Science Foundation Office of Polar Programs under Grant
No. 0338219, the University of Kansas, the University of Canterbury
Marsden Foundation, and the Cottrell Research Corporation for their
generous financial support. \message{Alexey Provorov and Igor
Zheleznykh (Moscow Institute of Nuclear Research, Moscow, Russia)
constructed the TEM horn antennas currently used as part of the
surface-noise veto. Derek Boyd performed important checks of the
overall system timing calibration. John Paden and Matt Peters provided
essential antenna expertise. George M.  Frichter, Adrienne Juett, Tim
Miller, Dave Schmitz, and Glenn Spiczak all performed essential work
in the initial construction phases of this experiment.} We also thank
the winterovers who staffed the experiment during the last six years
at the South Pole (Xinhua Bai, Allan Baker, Mike Boyce, Phil
Broughton, Marc Hellwig, Matthias Leupold, Karl Mueller, Michael
Offenbacher, Katherine Rawlins, Steffen Richter and Darryn Schneider),
as well as the excellent on-site support offered by Raytheon Polar
Services logistical personnel (particularly Rev. Al Baker, Jack
Corbin, Joe Crane and Paul Sullivan). Any opinions, findings, and
conclusions or recommendations expressed in this material are those of
the author(s) and do not necessarily reflect the views of the National
Science Foundation.  SR was partially supported by NSF grant AST
0307376.  We thank our RICE collaborators (particularly Shahid
Hussain, Doug McKay, John Ralston, David Seckel and Surujhdeo
Seunarine) for useful conversations and input.

\newpage

\section{Appendix - List of transients considered for matches}

Included below are transients with full spectral information.
Transient XRFs are insufficiently energetic for the current analysis.

%http://swift.gsfc.nasa.gov/docs/swift/archive/grb_table.html
SWIFT GRB / Time / Trigger  / RA /  Dec / Duration T90 /
Fluence (15-150 keV) / 1-sec Peak Photon Flux / Photon
Index / XRT Initial Temporal Decay Index / XRT Spectral Index
/ (Gamma) Redshift  / $<Livetime>$ / D1 threshold
%erg/cm2/s) (ph/cm2/sec) (PL = simple power-law, CPL = cutoff power-law)       

050908 /  05:42:31 / 154112 /  20.451 (01:21:48)  /  -12.962 (-12:57:45) /   
20 / 4.91 /  0.70 / 1.86 (PL)  / 1.33 / 3.9  / 3.340  / 0.92 / -0.389

050730  / 19:58:23 / 148225 / 212.063 (14:08:15.1)  / -3.740 (-03:44:24.0) / 
155    / 24.2 / 0.57 / 1.52 (PL)  / 2.3  / 1.8  / 3.97   / 0.75 / -0.370

050603 /  06:29:05 / 131560  / 39.982 (02:39:56) /   -25.195 (-25:11:41) /  
13    / 76.3 / 27.6 /  1.17 (PL) /  1.78 / 1.71 / 2.821 /  0.92 / -0.385

\vspace{1cm}

HETE GRB / Time  Class / Coords / Redshift / Epeak / T90 / 30-400 keV
fluence (erg/$cm^2$) / lightcurve / $<Livetime>$ / D1 threshold
%http://space.mit.edu/HETE/Bursts/Data/

%(note that XRFs are not considered for the calculations described above)

GRB 020124A / 10:41:15 / XRR /  09h 32m 49.0s /  -11d 27' 34'' /  3.2 /   
86.93 /  46.42 / 6.1e-06 /  0.83 / -0.50

GRB 020813A / 02:44:19 / GRB  / 19h / 46m 37.9s /  -19d 35' 16'' /  1.25 / 
142.1 /   89.29 / 8.4e-05 /  0.98 / -0.30

GRB 020903A / 10:05:37 / XRF /  22h 49m 01.0s /  -20d 55' 45'' /  0.25 /   
5 / 4.10 / 1.6e-08 /  0.92 / -0.22

GRB 030429A / 10:42:22 / XRF /  12h 13m 06.0s /  -20d 55' 59'' /  2.66  / 
35.04 /  10.30 / 3.8e-07 /  0.95 / -0.352

GRB 030723A / 06:28:17 / XRF /  21h 49m 29.7s /  -27d 42' 07'' /  2.3   / 
11.3  /   6.78 / 2.8e-07 /  0.96 / -0.352

\begin{table}%[hptb]
\caption{GRBs recorded in southern hemisphere, with incomplete spectral or
redshift information}
\begin{tabular}{c|c|c|c} \hline
GRB & Declination Angle  & $<Livetime>$ & Discriminator Threshold (V) 
\\ \hline
051111A & -2.64		 &  0.79   & -0.35 \\
051109B & -53.959	 &  0.946  & -0.48 \\
051021A & -45.519	 &  0.916  & -0.35 \\
051013A & -4.399	 &  0.886  & -0.4 \\
051007A & -0.512	 &  0.868  & -0.35 \\
051006B & -67.715	 &  0.958  & -0.45 \\
051004A & -10.869	 &  0.94   & -0.4 \\
051004B & -15.347	 &  0.946  & -0.4 \\
051001A & -30.4786 	 &  0.928  & -0.47 \\
050927A & -36.313	 &  0.934  & -0.44 \\
050921A & -37.034	 &  0.928  & -0.43 \\
050917A & -62.582	 &  0.97   & 0.55 \\
050917B & -21.242	 &  0.934  & -0.55 \\
050916A & -50.5703 	 &  0.976  & -0.57 \\
050915B & -66.5899 	 &  0.958  & -0.58 \\
050915A & -27.9835  	 &  0.97   & -0.58 \\
050911A & -37.1506 	 &  0.85   & -0.39 \\
050910A & -29.775	 &  0.928  & -0.45 \\
050908A & -11.0452 	 &  0.838  & -0.39 \\
050907A & -48.815	 &  0.898  & -0.39 \\
050830B & -45.23	 &  0.694  & -0.36 \\
050826A & -1.35683 	 &  0.856  & -0.41 \\
050822A & -45.9662 	 &  0.736  & -0.41 \\
050818A & -35.058	 &  0.97   & -0.51 \\
050817A & -24.93	 &  0.916  & -0.44 \\
050813A & -32.242	 &  0.718  & -0.4 \\
050813B & -60.716	 &  0.412  & -0.4 \\
050801A & -20.0719 	 &  0.886  & -0.38 \\
050728A & -26.984	 &  0.766  & -0.41 \\
050727A & -21.768	 &  0.922  & -0.41 \\
050726A & -31.9359 	 &  0.952  & -0.41 \\
052726B & -22.655	 &  0.904  & -0.41 \\
050725A & -16.466	 &  0.898  & -0.42 \\
050724A & -26.459  	 &  0.904  & -0.40 \\
050721A & -27.6189 	 &  0.454  & -0.29 \\
050721B & -1.44		 &  0.826  & -0.29 \\
050715B & -0.059	 &  0.514  & -0.34 \\
050714A & -15.528	 &  0.754  & -0.38 \\
050713A & -12.264	 &  0.604  & -0.33 \\
\end{tabular}
\end{table}

\begin{center}
\begin{tabular}{c|c|c|c}
050709A & -37.0224 	 &  0.808  & -0.31 \\
050703A & -43.059  	 &  0.97   & -0.51 \\
050701A & -58.5844 	 &  0.478  & -0.38 \\
050701B & -1.4526	 &  0.616  & -0.47 \\
050626A & -63.1342 	 &  0.388  & -0.37 \\
050623A & -27.331	 &  0.838  & -0.46 \\
050614B & -42.492	 &  0.874  & -0.29 \\
050612A & -63.6821 	 &  0.658  & -0.38 \\
050603A & -24.8182 	 &  0.88   & -0.39 \\
050602A & -20.198	 &  0.874  & -0.37 \\
050509C & -43.1655 	 &  0.736  & -0.45 \\
050418A & -18.538	 &  0.952  & -0.40 \\
050416A & -27.907	 &  0.994  & -0.23 \\
050412A & -0.799	 &  0.688  & -0.42 \\
050412B & -1.252	 &  0.688  & -0.42 \\
050406A & -49.8125 	 &  0.934  & -0.30 \\
050402A & -20.422	 &  0.802  & -0.35 \\
050331A & -40.5617 	 &  0.904  & -0.29 \\
050331B & -42.678	 &  0.928  & -0.29 \\
050326A & -70.6288 	 &  0.934  & -0.42 \\
050326B & -71.375	 &  0.922  & -0.42 \\
050323A & -40.582	 &  0.844  & -0.27 \\
050323B & -20.417	 &  0.916  & -0.32 \\
050322A & -42.093	 &  0.832  & -0.32 \\
050318A & -45.6045 	 &  0.868  & -0.24 \\
050226A & -5.354	 &  0.982  & -0.55 \\
050223A & -61.5275 	 &  0.904  & -0.18 \\
050222A & -53.777	 &  0.856  & -0.16 \\
050219B & -56.2424 	 &  0.934  & -0.28 \\
050219A & -39.3172 	 &  0.904  & -0.28 \\
050217A & -43.036	 &  0.964  & -0.48 \\
050213A & -53.791	 &  0.91   & -0.78 \\
050211A & -20.41	 &  0.856  & -0.46 \\
050203A & -52.384	 &  0.43   & -0.50 \\
050202A & -38.72	 &  0.266  & -0.67 \\
050130A & -20.405	 &  0.85   & -0.23 \\
050129A & -43.013	 &  0.598  & -0.21 \\
050121A & -20.387	 &  0.01   & -0.59 \\
050121B & -37.84	 &  0.976  & -0.59 \\
050108A & -37.848	 &  0.04   & -0.18 \\
\end{tabular}
\end{center}

\begin{center}
\begin{tabular}{c|c|c|c}
050107A & -20.409	 &  0.97   & -0.56 \\
050107B & -15.637	 &  0.892  & -0.56 \\
041224A & -5.344	 &  0.97   & -0.88 \\
041223A & -36.9271 	 &  0.322  & -0.78 \\
040730A & -55.5492 	 &  0.922  & -0.17 \\
040330A & -20.4104 	 &  0.91   & -0.33 \\
040309A & -20.3924 	 &  0.832  & -0.30 \\
040228A & -16.612  	 &  0.892  & -0.37 \\
040223A & -40.0667 	 &  0.244  & -0.34 \\
031015A & -20.525	 &  0.976  & -0.65 \\
030821A & -44.6431 	 &  0.988  & -0.87 \\
030725A & -49.3178 	 &  0.982  & -0.35 \\
030723A & -26.2868 	 &  0.92   & -0.28 \\
030716A & -67.8673 	 &  0.934  & -0.28 \\
030706A & -47.4994 	 &  0.976  & -0.35 \\
030629A & -42.571	 &  0.976  & -0.28 \\
030528A & -21.3806 	 &  0.964  & -0.29 \\
030429A & -19.0862 	 &  0.94   & -0.35 \\
030418A & -6.972	 &  0.976  & -0.35 \\
030323A & -20.2297 	 &  0.592  & -0.62 \\
021021A & -0.38333 	 &  0.868  & -0.35 \\ 
020903A & -19.0708 	 &  0.892  & -0.22 \\
020813A & -18.3987 	 &  0.976  & -0.30 \\
020812A & -4.70056 	 &  0.952  & -0.24 \\
020801A & -52.2297 	 &  0.952  & -0.35 \\
020629B & -17.2772 	 &  0.976  & -0.35 \\
020305A & -13.6967 	 &  0.982  & -0.50 \\
020124A & -10.5406 	 &  0.52   & -0.50 \\
\end{tabular}
\end{center}

\end{document}